\documentclass[aps,prd,groupedaddress,floatfix,longbibliography,superscriptaddress,nofootinbib]{revtex4-2}

\usepackage{amsmath}
\usepackage{amssymb}
\usepackage{amsfonts}
\usepackage{graphicx}
\usepackage{bbold}
\usepackage{bm}
\usepackage[usenames,dvipsnames,svgnames]{xcolor}
\usepackage{hyperref}
\hypersetup{colorlinks=true, linkcolor=Blue, citecolor=Blue,urlcolor=Blue}
\usepackage{color,epsfig}
\usepackage{physics}
\begin{document}

%\title{Magnetic moment contributions to the planar Hall effect in Weyl semimetals \\ Strain-induced planar Hall effect in Weyl semimetals \\ Planar Hall effect in Weyl semimetals induced by pseudoelectromagnetic fields}

\title{Emergent universality of free fall from quantum mechanics}

\author{Juan A. Cañas}
\email{juan.canas@correo.nucleares.unam.mx}
\address{Instituto de Ciencias Nucleares, Universidad Nacional Aut\'{o}noma de M\'{e}xico, 04510 Ciudad de M\'{e}xico, M\'{e}xico}

\author{A. Mart\'{i}n-Ruiz}
\email{alberto.martin@nucleares.unam.mx}
\address{Instituto de Ciencias Nucleares, Universidad Nacional Aut\'{o}noma de M\'{e}xico, 04510 Ciudad de M\'{e}xico, M\'{e}xico}

\author{J. Bernal}
\address{Divisi\'on Acad\'emica de Ciencias B\'asicas, Universidad Ju\'arez Aut\'onoma de Tabasco, 86690, Cunduac\'an, Tabasco, M\'exico}

\begin{abstract}
Classical and quantum mechanical descriptions of motion are fundamentally different. The universality of free fall (UFF) is a distinguishing feature of the classical motion (which has been verified with astonishing precision), while quantum theory tell us only about probabilities and uncertainties thus breaking the UFF. There are strong reasons to believe that the classical description must emerge, under plausible hypothesis, from quantum mechanics. In this Essay we show that the UFF is an emergent phenomenon: the coarse-grained quantum distribution for high energy levels leads to the classical distribution as the lowest order plus quantum corrections. We estimate the size of these corrections on the Eötvös parameter and discuss the physical implications.
\end{abstract}

\maketitle

\begin{center}
{\it{   Essay written for the Gravity Research Foundation 2024 Awards for Essays on Gravitation}}
\end{center}

%-----------------------------------------------------------

\section{Introduction}

Legend has it that Galileo dropped two spheres of different masses from the top of the Leaning Tower of Pisa to demonstrate that the acceleration at which objects fall is the same irrespective of their mass, establishing the so called universality of free fall (UFF), often referred as the weak equivalence principle. The advent of Newton's mechanics made clear that Galileo's principle was due to the equivalence between inertial and gravitational masses. In 1915, Einstein extended this postulate into the equivalence principle, one of the cornerstones of general relativity. So far, several tests have been performed with pendula or torsion balances leading to extremely accurate confirmation of the equality of gravitational and inertial masses at the classical level \cite{PhysRevLett.93.261101, PhysRevLett.100.041101}. Gravity-induced interference experiments have also confirmed this result in the quantum domain \cite{PhysRevLett.34.1472}. 

UFF is exactly true for classical point particles in Newtonian gravity and general relativity; however as Visser pointed out, classical structured particles (e.g. with a mass quadrupole moment) need not obey the UFF \cite{doi:10.1142/S0218271817430271}. The case is even more subtle in quantum theory. As we know, in classical mechanics it is possible to know the exact position and momentum of a particle at any given time. However, the conceptual framework of the quantum theory is fundamentally different and only tell us about the probability of finding a particle at certain position. Therefore, the Galilean form of the UFF, basically that all bodies follow the same trajectory when freely falling in the gravity field with the same initial conditions, is inappropriate to be applied to quantum probes. Indeed, one would expects violations to the UFF with quantum probes since the dynamics is mass-dependent, i.e. mass does not cancel out from the Schrödinger equation for a particle in a uniform gravity field. In this framework, the nonuniversality of the quantum free fall has been investigated by using sharply peaked wave packets \cite{doi:10.1142/S0218271817430271}, Schrödinger cat states \cite{PhysRevD.55.455}, Gaussian \cite{Ali_2006} and nonGaussian \cite{Chowdhury_2012} wave packets and the cut-the-wave procedure \cite{WepDiff2022}. 

The failure of the Galilean statement of the weak equivalence principle at quantum level have prompted the search for a more fundamental statement that is directly applied to subatomic particles by which it is expected to search for the clues of constructing a quantum theory of gravity \cite{Zych_2018}. On a deeper level, Padmanabhan postulated the existence of certain unknown microscopic degrees of freedom of the spacetime such that when coarse-grained (in the long wavelength limit), lead to Einstein gravity as the lowest order with the Lanczos-Lovelock type quantum corrections \cite{PADMANABHAN_2008}. In this Essay we adopt a simplest framework by assuming the universality of quantum mechanics and the validity of the UFF at the classical level. We aim to explore how coarse-graining ideas lead to the emergence of the weak equivalence principle as the lowest order when quantum mechanics is applied to a freely falling macroscopic particle which resides in an energy eigenstate. We estimate the size of quantum corrections on the Eötvös parameter.

%------------------------------------------------------

\section{Locally averaging quantum mechanics}

It is well-known that Einstein firmly believed that quantum mechanics was incomplete, a fact reflected in his famous quote {\it{God does not play dice with the universe}}. He was convinced that the spreading of wavepackets and collapsing wave function provided an inadequate and incomplete description of reality. This subtlety was discussed by Einstein and Born in their famous letters \cite{einstein1971born}. To illustrate his concerns, Einstein invoked the problem of a macroscopic particle moving freely in a one-dimensional box and ask himself what quantum mechanics can tell us about the motion of this particle residing in an energy eigenstate. Taking a sufficiently large principal quantum number (such that the de Broglie wavelength is small as compared with the size of the box), Einstein realized that the coarse-grained probability density is uniform inside the box, as expected from the classical ensemble theory.  The emergence of coarse-grained descriptions from more fundamental theories is beyond quantum mechanics, ranging from thermodynamics to quantum field theory. What Einstein disliked is that quantum mechanics requires to speak of an ensemble and tells us nothing about our individual particle. Following this line of reasoning, in this Essay we discuss the coarse-graining emergence of the UFF in the quantum domain. To this end some technicalities are in order.

Our starting point is the usual manner in which the quantum-classical correspondence is discussed for periodic systems: a direct comparison between the quantum and classical probability densities \cite{Yoder_2006} (as Einstein did in his debate with Born) to show their convergence when the former is coarse-grained. The quantum probability density is defined by $\rho ^{\mbox{\scriptsize qm}} _{n} (x) = \vert \psi _{n} (x) \vert ^{2}$, where $\psi _{n} (x)$ is an energy eigenstate of the system, and the classical distribution is properly defined as  $\rho _{\mbox{\scriptsize cl}} (x) = \frac{2}{\tau \vert v (x) \vert }  $, where $\tau$ and $v(x)$ are the classical period and local velocity, respectively. Intuitively, $\rho _{\mbox{\scriptsize cl}} (x)$ is derived by the argument of how much time spent the particle in a given interval. A direct comparison between these distributions is not satisfactory at all, since in the limit of large quantum number $n$, the highly oscillatory quantum distribution $\rho ^{\mbox{\scriptsize qm}} _{n} (x)$ does not converge pointwise to the smooth classical distribution $\rho _{\mbox{\scriptsize cl}} (x)$ \cite{Robinett_1995,Yoder_2006}, as Einstein pointed out. Instead, the connection should be understood in a distributional sense. The convergence in distribution means that the quantum distribution $\rho ^{\mbox{\scriptsize qm}} _{n} (x)$, if averaged locally in a finite interval (coarse-grained), approaches to the classical distribution $\rho _{\mbox{\scriptsize cl}} (x)$ for a sufficiently large quantum number $n$. This fact is expressed by
\begin{align}
	\rho _{\mbox{\scriptsize cl}} (x) = \lim _{n \gg 1} \frac{1}{2 \epsilon _{n} } \int _{x - \epsilon _{n}} ^{x + \epsilon _{n}} \rho ^{\mbox{\scriptsize qm}} _{n} (x') \, dx' , \label{Local_Average}
\end{align}
where the interval $\epsilon _{n}$ decreases as increasing the quantum number $n$ \cite{Liboff_1984}. Now we reformulate the local averaging process (\ref{Local_Average}) by assuming that (i) the classical motion is periodic (i.e. the particle bounces back and forth between the turning points) and (ii) the quantum distribution rapidly oscillates for large quantum numbers \cite{Canas2022}. These natural assumptions imply that both, the classical and quantum probability distributions can be expressed as Fourier expansions with coefficients $\varrho _{\mbox{\scriptsize cl}} (p)$ and $\varrho ^{\mbox{\scriptsize qm}} _{n} (p)$, respectively. Inserting the expansions into Eq. (\ref{Local_Average}) leads to the conclusion that the quantum coefficient $\varrho ^{\mbox{\scriptsize qm}} _{n} (p)$ approaches, asymptotically, to the classical coefficient $\varrho _{\mbox{\scriptsize cl}} (p)$ for $n\gg 1$, i.e.
\begin{align}
	\varrho ^{\mbox{\scriptsize qm}} _{n} (p) \sim \varrho _{\mbox{\scriptsize cl}} (p) + \mathcal{O} (1/n) .  \label{Asymptotic_Fourier}
\end{align}
This result says that the resulting distribution in coordinate representation is the classical probability density $\rho _{\mbox{\scriptsize cl}} (x)$ at the lowest order with quantum corrections decaying as a power of $1/n$ \cite{Bernal_2013, Mart_n_Ruiz_2014, Mart_n_Ruiz_2013}. There are two subtle points that deserves discussion.

Above we have taken the limit $n \gg 1$, but how large the principal quantum number should be? Physically, we are describing the physics of a macroscopic particle (with energy $E _{\mbox{\scriptsize cl}}$) residing in the $n$th energy eigenstate (with energy $E _{n}$). Therefore, $n$ should be determined by equating $E _{\mbox{\scriptsize cl}} = E _{n}$. For a free particle in a box, for example, this expression yields a precise value $n = \sqrt{E _{\mbox{\scriptsize cl}} / E _{1} } \gg 1$, where $E _{1}$ is the ground state. In the case of the harmonic oscillator, the quantum probability density confines progressively as the principal quantum number is increased. The turning points, which correspond to the maximum amplitude of the macroscopic oscillator, are determined by equating $E _{\mbox{\scriptsize cl}} = E _{n}$ and hence the  value of $n$ is fixed. These examples tell us that $n$ is large but finite. Therefore, the additional term appearing in Eq. (\ref{Asymptotic_Fourier}) is  finite and can be interpreted as a small quantum correction. But, what is the physical origin of such correction term? To answer this question, we recall that the macroscopic behavior was attained by assuming a sufficiently large quantum number, but we kept fixed $\hbar$. In our view, $\hbar$ is a fundamental constant of nature whose numerical value although small is nonzero. Indeed, resorting to the Bohr-Sommerfeld quantization rule, the correction term in powers of $1/n$ corresponds to powers of $\hbar / S$, where $S$ is the classical action. Therefore, the corrections terms are remnants of the underlying quantum theory at the macroscopic level.

%%---------------------------------%%%%%%%%%%%%%

\section{Bouncing particles off the floor}

Let us consider a particle of mass $m$ that is initially at rest and dropped from a height $h$ above a reflecting surface at $z=0$. In the following discussion we assume the equivalence between inertial and gravitational masses. The direction of acceleration due to gravity is always vertically downward. The classical probability density is found to be
\begin{align}
	\rho _{\mbox{\scriptsize cl}} (z) = \frac{ H (h - z) }{2\sqrt{h\left(h-z\right)}} , \label{Classical_dist}
\end{align}
where $H (x)$ is the Heaviside step function. Notice that this expression does not depend on neither the gravitational acceleration nor the mass, thus reflecting the classical UFF. Interestingly, it diverges near to $z=h$, where the particle slows down and reverses direction, but it is finite at $z=0$, where the particle bounces.

The quantum-mechanical problem is simple as well. The solution to the stationary Schr\"{o}dinger equation which properly decays for $z \to \infty$ and vanishes on the floor $\psi _{n} (z= 0) = 0$, leads to the quantum probability density
\begin{align}
	\rho ^{\mbox{\scriptsize qm}} _{n} (z) = \left[  \frac{1}{ \sqrt{l _{g}} } \frac{\mbox{Ai} (a _{n} + z / l _{g}) }{\mbox{Ai} ' (a _{n})} \right] ^{2} H (z) ,  \label{Quantum_dist}
\end{align}
where $\mbox{Ai} (x)$ is the Airy function, $a_{n}$ is its $n$-th zero and $l _{g} = \left( \frac{\hbar ^{2}}{2m^{2}g} \right) ^{1/3} $ is the gravitational length. The resulting quantized energies are $E _{n} = - mgl _{g} a _{n}$ with $n=1,2, \cdots$, and the corresponding classically allowed heights are $h _{n} = E_{n} /(mg) = - a _{n} l _{g} $. As soon as such a height $h _{n}$ is reached, the quantum distribution starts approaching zero exponentially fast, besides it depends on both the mass $m$ and the gravitational acceleration $g$. The GRANIT experiment has confirmed that a noncoherent beam of ultracold neutrons propagating upwards in the Earth's gravity reaches precisely these quantized heights \cite{Nesvizhevsky_2002}.

\begin{figure}
	\centering
	\includegraphics[scale =0.4]{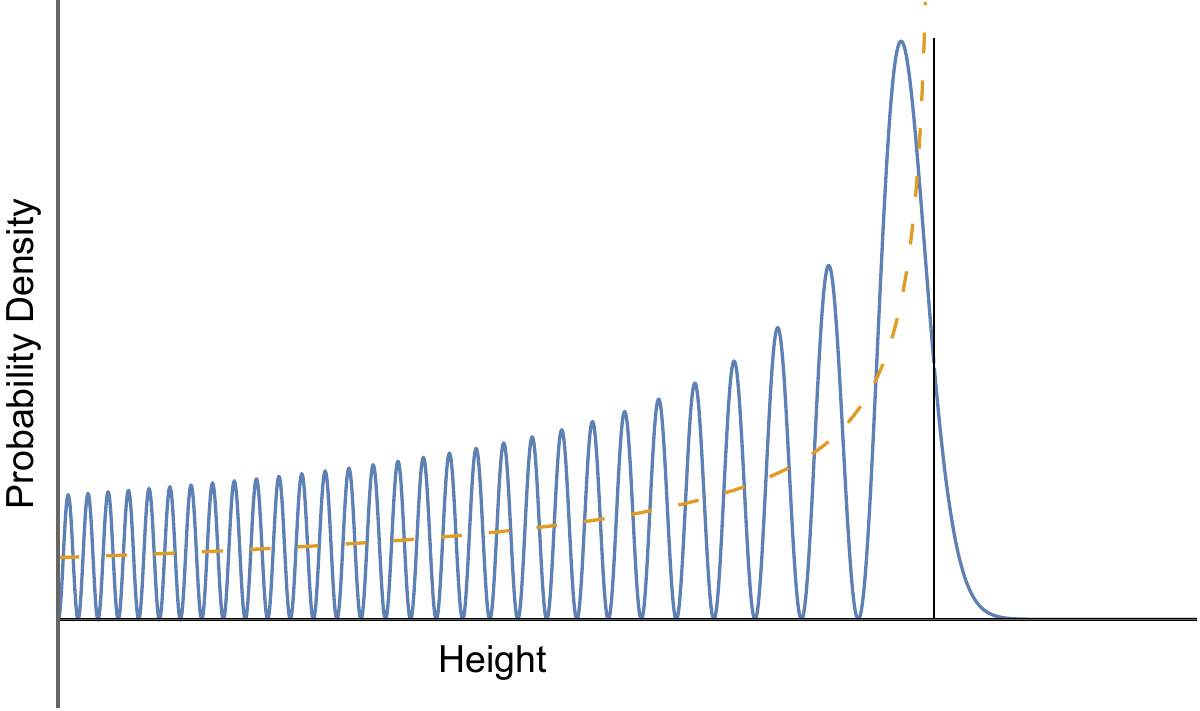}
	\includegraphics[scale =0.4]{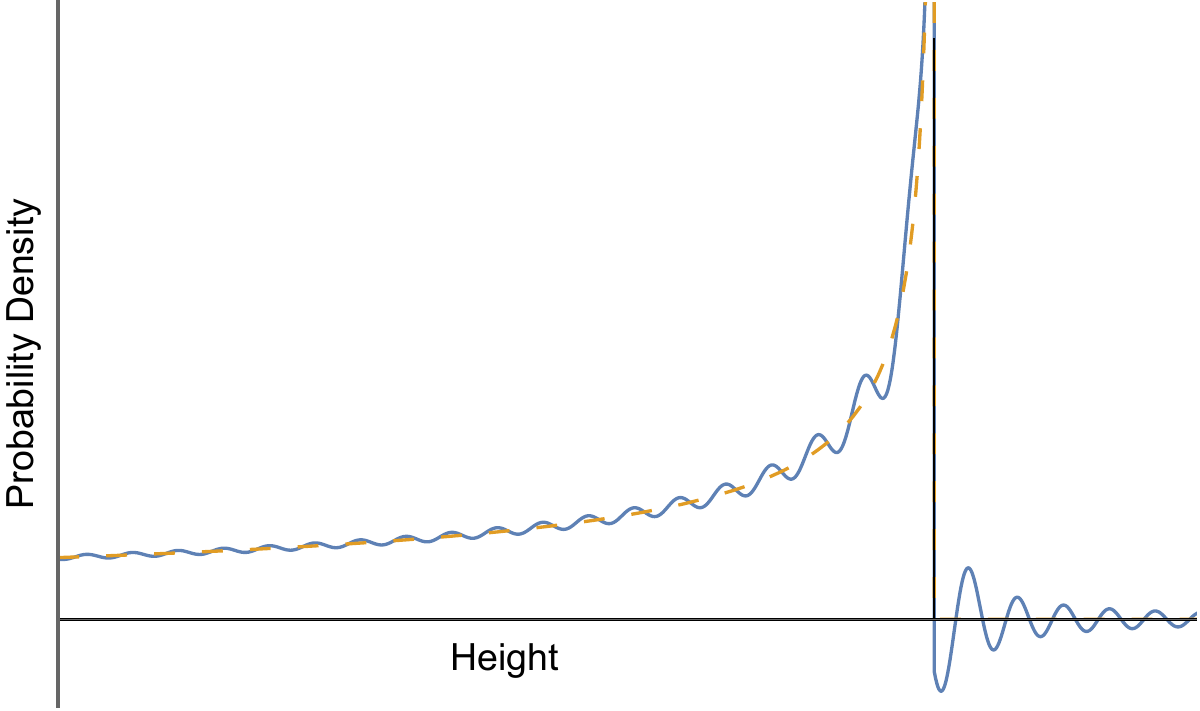}
	\caption{ The left panel sketches the classical (dashed-yellow line) and quantum (continuous-blue line) probability densities as a function of the height. The right panel depicts the classical (dashed-yellow line) and the asymptotic quantum (continuous-blue line) probability densities as a function of the height. The vertical line marks the turning point.}
	\label{Figure_Essay}
\end{figure}

In order to illustrate visually that the coarse-grained quantum probability density  yields the classical distribution for $n \gg 1$, in the left panel of figure  \ref{Figure_Essay} we sketch the classical (yellow-dashed line) and quantum (blue-continuous line) distributions as a function of the height for $n=30$. We now use the theory of the previous section to evince the emergence of the macroscopic behavior. To this end, we have to compute first the Fourier coefficient 
\begin{align}
	\varrho ^{\mbox{\scriptsize qm}} _{n} (p) =    \frac{1 }{ {l _{g}} \, \mbox{Ai} ^{\prime \, 2} (a _{n})} \int _{0} ^{\infty} \mbox{Ai} ^{2} (a _{n} + z / l _{g}) \, e^{-ipz/\hbar } dz
\end{align}
and next evaluate its asymptotic behavior for $n \gg 1$. The required integral (to the best of our knowledge) cannot be evaluated in a closed form. However, for our purposes in this Essay, we only need the asymptotic expansion of the integral. To do this we use the method of Albright \cite{doi:10.1142/p709}, since some basic primitives of the Airy functions allow us to express such integral as a series in inverse powers of $a _{n}$. Leaving aside the technical details\footnote{To evaluate the asymptotics of the integral $\int _{0} ^{\infty} \mbox{Ai} ^{2} (a _{n} + z / l _{g}) \, e^{-ipz/\hbar } dz$ we first change variable to $x = a _{n} + z / l _{g}$ and then Taylor expand the exponential $e^{-ipx/\hbar } = \sum _{k=0} ^{\infty} \frac{1}{k!} (-ipx/\hbar) ^{k} $. The resulting integrals $\int _{a _{n}} ^{\infty} x ^{k} \, \mbox{Ai} ^{2} (x) \, dx$ can be evaluated by using the method of Albright. Next, we cut the series up to the desired order in $1/a _{n}$ (since we take $n \gg 1$). Here we retain the two first leading terms. Finally, we evaluate the summations over $k$.} and keeping only the first two leading terms, the asymptotics of the Fourier coefficient can be written as $\varrho ^{\mbox{\scriptsize qm}} _{n} (p) \approx  \varrho ^{\mbox{\scriptsize qm} (0)} _{n} (p) + \varrho ^{\mbox{\scriptsize qm} (1)} _{n} (p)$, where
\begin{align}
	\varrho ^{\mbox{\scriptsize qm} (0)} _{n} (p) &= - \frac{1  }{  \sqrt{ i Q _{n} }} \, \mathcal{F} ( \sqrt{ i Q _{n} } )  , \label{Leading_FC} \\ \varrho ^{\mbox{\scriptsize qm} (1)} _{n} (p) &= \frac{1}{96 a _{n} ^{3}}  \left[  -  \left(15-8 i Q _{n} ^{3} \right) \varrho ^{\mbox{\scriptsize qm} (0)} _{n} (p) + 2 Q _{n} (2 Q _{n} + 5 i) - 15 \right] , \label{Subleading_FC}
\end{align}
where $Q _{n} = p h _{n} / \hbar$ and $\mathcal{F}$ is the Dawson function, which is defined in terms of the imaginary error function by $\mathcal{F} (x) = \frac{\sqrt{\pi}}{2} e ^{-x^{2}} \mbox{erfi} (x)$ \cite{Canas2022}. In coordinate representation, the probability density can be written as the sum of two terms as well: $\rho ^{\mbox{\scriptsize qm}} _{n} (z) \approx  \rho ^{\mbox{\scriptsize qm} (0)} _{n} (z) + \rho ^{\mbox{\scriptsize qm} (1)} _{n} (z)$. These are obtained by inverse Fourier transforming the coefficients (\ref{Leading_FC}) and (\ref{Subleading_FC}). For the leading term one gets
\begin{align}
	\rho ^{\mbox{\scriptsize qm} (0)} _{n} (z) = \frac{1}{2 \pi \hbar } \int \varrho ^{\mbox{\scriptsize qm} (0)} _{n} (p) \, e ^{i pz/ \hbar} \, dp = \frac{ H (h _{n} - z) }{2\sqrt{h _{n} \left(h _{n} -z\right)}}  ,
\end{align}
which is exactly the classical probability density $\rho _{\mbox{\scriptsize cl}} (z)$, given by Eq. (\ref{Classical_dist}), with the classical height $h$ replaced by the quantum-mechanical turning point $h _{n}$. The subleading term $\rho ^{\mbox{\scriptsize qm} (1)} _{n} (z) = \frac{1}{2 \pi \hbar } \int \varrho ^{\mbox{\scriptsize qm} (1)} _{n} (p) \, e ^{i pz/ \hbar} \, dp $ cannot be expressed in terms of simple functions, but it can be evaluated numerically. In the right panel of figure \ref{Figure_Essay} we sketch the classical (yellow-dashed line) versus the coarse-grained quantum (blue-continuous line) distributions as a function of the height for $n=100$. This figure reveals that the correction term $\rho ^{\mbox{\scriptsize qm} (1)} _{n} (z)$ consists of small oscillations around the classical distribution (for $0<z<h _{n}$) which we interpret as a residual quantum mechanical behavior at the macroscopic level. Besides, we observe that the probability density remains nonzero beyond the turning point ($z>h _{n}$) thus implying a non-Newtonian behavior. Finally, having showed that the quantum probability density converges in a distributional sense to its classical counterpart for $n \gg 1$,  it is clear that the expectation value of any position-dependent observable $f$ correctly yields the classical result as the lowest order, i.e. $\left< f(z) \right> _{\mbox{\scriptsize qm}} \approx \left< f(z) \right> _{\mbox{\scriptsize cl}}$. As larger $n$ is, better is the convergence.

\section{Quantum corrections to the Eötvös parameter}

Modern tests of the UFF generally involve measuring the relative acceleration between two different test bodies freely falling under the gravity field. Such tests are characterized by the Eötvös parameter:
\begin{align}
	\eta = 2 \frac{a _{1} - a _{2}}{a _{1} + a _{2}} ,  \label{Eotvos}
\end{align}
where $a _{i}$ is the gravitational acceleration of object $i$ with respect to the source mass. The validity of the UFF with classical probes have been established with astonishing precision, reaching uncertainties at the level of $10^{-13}$ with torsion balances \cite{PhysRevLett.93.261101, PhysRevLett.100.041101} and $2 \times 10^{-14}$ with accelerometers onboard the Microscope satellite \cite{PhysRevLett.119.231101} in measuring $\eta$. In the quantum domain, the UFF has been confirmed up to an uncertainty of $10 ^{-7}$ by using atomic interferometry between free-falling correlated atoms  \cite{Albers_2020}. The large gap between these measurements opens the possibility for deviations to the UFF due to quantum effects, which nevertheless are undetectable with the current experimental precision. As a consequence of the remnant quantum behavior at the macroscopic level predicted in the previous Section for a particle freely-falling in the Newtonian gravity field, we can give an estimation of quantum corrections to the Eötvös parameter (\ref{Eotvos}), as we show in the following.

To this end we have to evaluate the gravitational force upon an extended object. As we now, for a classical point particle the Newtonian gravitational force is ${\bf{F}} = - m  \nabla  \phi ({\bf{r}})$. As a natural assumption, the net force can be obtained by integrating over the mass distribution as described by the probability density $\rho ({\bf{r}})$, i.e. ${\bf{F}} = - m  \int \rho ({\bf{r}}) \, \nabla  \phi ({\bf{r}}) \, d ^{3} {\bf{r}}$. Using this expression together with a multipolar expansion of the gravitational potential $\phi ({\bf{r}})$, Visser derived an effective acceleration in terms of probability quadrupole moments of wavepackets \cite{doi:10.1142/S0218271817430271}. In the problem at hand, however, we keep the usual Newtonian gravitational potential and derive an effective acceleration due to the quantum corrections discussed in the previous section, i.e. $a _{\mbox{\scriptsize eff}} = \int \left[  \rho ^{\mbox{\scriptsize qm} (0)} _{n} (z) + \rho ^{\mbox{\scriptsize qm} (1)} _{n} (z) \right] g \, dz $. A simple calculation\footnote{The expression we compute is $\left<  z ^{k} \right>  ^{\mbox{\scriptsize qm}} _{n} \approx \int _{0} ^{\infty} z ^{k} \, [ \, \rho ^{ (0) } _{n} (z)  + \rho ^{ (1) } _{n} (z) \, ] \, dz$ and next take $k=0$. The former term yields the classical average value $\left<  z ^{k} \right>  _{\mbox{\scriptsize cl}}$ and the latter can be evaluated by using the standard procedure to regularize infinite trigonometric integrals, i.e. $\int _{0} ^{\infty} z ^{k} \,   \rho ^{ (1) } _{n} (z) \, dz =  \frac{1}{2 \pi \hbar } \int dq \,  \rho ^{ (1) } _{n} (q)  \lim _{\epsilon \to 0 ^{+}} \int _{0} ^{\infty} z ^{k} \,   e ^{i qz/ \hbar } e ^{- \epsilon z} \, dz  = \ \frac{1}{2 \pi \hbar } \int  \rho ^{ (1) } _{n} (q)  \frac{\Gamma (k+1)}{(-i q / \hbar ) ^{k+1}} dq = \left<  z ^{k} \right>  _{\mbox{\scriptsize cl}} \,  \frac{-4 k^3+6 k^2+k+6}{48 \, a _{n} ^{3}}$.} yields $a _{\mbox{\scriptsize eff}} = g \left( 1 + \Delta _{n} \right)$, where $\Delta _{n} = 1/ (8 a _{n} ^{3})$.

For reference, we take the experimental configuration of Ref. \cite{Albers_2020} where ${}^{87}$Rb and ${}^{39}$K atoms fall in the presence of the gravity field from a macroscopic height of $h = 19$ cm (200 ms). This experiment yields a determination of the  Eötvös parameter of $-1.9 \times 10 ^{-7}$. In our framework, since masses of these atoms are different, the corresponding quantum numbers are too and hence the Eötvös parameter (\ref{Eotvos}) acquires the quantum correction $\delta \eta = \Delta _{n _{\mbox{\scriptsize Rb}}} - \Delta _{n _{\mbox{\scriptsize K}}}$, where $\Delta _{n _{\mbox{\scriptsize Rb}}} = h ^{3} / (8 l _{g \, \mbox{\scriptsize Rb}} ^{3} ) $ for rubidium and similarly for the potassium, being $l _{g}$ the corresponding gravitational length. A simple estimation produces $\delta \eta = - 1.97 \times 10 ^{-18}$, which is too far from the current experimental precision and hence undetectable for the time being. 

%%%%----------------------------------------------------------------------------%%%%%%%

\section{Discussion}

The UFF is a distinguishing feature of classical point particles when freely falling in the gravitational field. This universality is naturally broken in the quantum domain. The emergence of the UFF from quantum mechanics is subtle, since the conceptual frameworks of both theories are fundamentally different. In this Essay we have introduced a simple method to compute in an analytical fashion the coarse-grained quantum probability distribution, such that the classical distribution emerges at the lowest order with quantum corrections. Using this method, we show that the UFF is an emergent phenomenon: the coarse-grained quantum distribution for a particle in the presence of the gravity field and residing in a high-energy eigenstate ($n \gg 1$) leads to the classical distribution as the lowest order plus quantum corrections. We evaluated  modifications to the Eötvös parameter due to the remnant quantum corrections at the macroscopic level. While certainly challenging, experimentally probing the ideas developed in this Essay is not entirely implausible.

%{\color{blue}{\it{Acknowledgements. }}}{L.M.O. was supported by the CONACyT PhD fellowship No. 834773. A.M.-R. has been partially supported by DGAPA-UNAM Project No. IA102722 and by Project CONACyT (M\'{e}xico) No. 428214}

\bibliography{references.bib}
\end{document}